\documentstyle[twoside,fleqn,espcrc2,epsf,amssymb]{article}

\newcommand{\AmS}{{\protect\the\textfont2
  A\kern-.1667em\lower.5ex\hbox{M}\kern-.125emS}}

\newcommand{\intx}{\int d^4 x\,}
\newcommand{\be}{\begin{equation}}
\newcommand{\ee}{\end{equation}}
\newcommand{\bea}{\begin{eqnarray}}
\newcommand{\eea}{\end{eqnarray}}
\newcommand{\kp}{\kappa}
\newcommand{\dl}{\delta}
\newcommand{\Lm}{\Lambda}
\newcommand{\rh}{\rho}
\newcommand{\ta}{\tau}
\newcommand{\ra}{\rightarrow}
\newcommand{\Om}{\Omega}
\newcommand{\half}{\frac{1}{2}}
\newcommand{\dmu}{\partial_{\mu}}
\newcommand{\dnu}{\partial_{\nu}}
\newcommand{\zt}{\zeta}
\newcommand{\ph}{\phi}
\newcommand{\lm}{\lambda}
\title{Remarks on the quantum gravity interpretation of 4D dynamical
triangulation}

\author{Jan Smit\address{Institute for Theoretical Physics,
        University of Amsterdam, \\
        Valckenierstraat 65, 1018 XE Amsterdam,
        The Netherlands}
}

\begin{document}

\begin{abstract}
We review some of the phenomenology in 4D dynamical
triangulation and explore its interpretation in terms 
of a euclidean effective action of the continuum form
$\intx \sqrt{g}\, [\mu -\frac{1}{16\pi G}\, R + \cdots]$.
\end{abstract}

\maketitle

4D Dynamical Triangulation (DT) is a formulation
of purely euclidean geometrodynamics \cite{Jo96}. Its
customary canonical partition function reads
\be
Z(\kp_2,N) = \sum_{{\cal T}, N_4 = N} e^{\kp_2 N_2},
\ee
where the summation is over all distinct triangulations
${\cal T}$ (describing simplicial manifolds)
with $S^4$ topology and $N_k$, $k=0,1, \ldots,4$
is the number of $k$-simplices.
The way the model is derived shows that $\kp_2 \propto
1/G_0$, with $G_0$ the bare Newton constant. In this talk I
explore a tentative continuum interpretation of the model,
described by the effective theory
\bea
Z &=& \int_{-i\infty}^{i\infty} d\mu \int Dg\, e^{-S_{\rm
eff}[g] + \mu V},\\
S_{\rm eff} [g] &=& \int d^4 x\, \sqrt{g}\, \left( \mu -
\frac{R}{16\pi G}
+ \cdots\right).
\label{effact}
\eea
Here the path integral is over real metrics modulo
coordinate transformations,  $G$ denotes a renormalized
Newton constant and the $\cdots$ indicate higher
derivative terms like $R^2$, etc. There may also be nonlocal terms related
to the conformal anomaly \cite{AnMo92}. The integral over
$\mu$ produces the volume fixing delta function $\dl(\intx
\sqrt{g} -V)$. If this integral were done in the saddle
point approximation, the saddle point value $\mu_c$ would be
related to a renormalized cosmological constant by 
\be
\mu_c = \frac{\Lm}{8\pi G}.
\label{cosco}
\ee

In general the higher order terms may be present in an
effective action. Here we
expect them to regulate the unboundedness of the Einstein-
Hilbert part in the ultraviolet. The underlying DT model is
finite and the nontrivial results of numerical simulations show that the
action is not stuck at its minimum value in the relevant
$\kp_2$ range. So entropy effects somehow provide a
regulating effect which can be implemented in $S_{\rm eff}$
through the higher order terms.

Let us now go through some of the DT phenomenology found in
numerical simulations and compare this with the effective
action (\ref{effact}).\\
{\bf 1}. There is a crumpled phase ($\kp_2 < \kp_2^c$) and an
elongated phase ($\kp_2 > \kp_2^c$),
which has characteristics of a branched polymer \cite{polymer}.
Since $\kp_2 \propto 1/G_0$ this suggests that the effective theory
also has a transition at some value of $1/G$, presumably near $1/G=0$. \\
{\bf 2}. Recent evidence 
\cite{firsto} indicates strongly that the
transition between the two phases is first order. This is
often seen as a disaster for a continuum interpretation.
However, we have suggested earlier \cite{BaSm95} that continuum
behavior may be automatic without tuning to a second order
phase transition. This happens in DT in two dimensions. A
field theoretic example is provided by the purely discrete
$Z(n)$ gauge-Higgs systems. For sufficiently large but
finite $n$ these models have a Coulomb phase with
massless photons and a Higgs phase separated by a first
order transition 
\cite{Zn}. The models can
approximate the continuum abelian Higgs model arbitrarily
well. Analysis of the continuum model shows the
possibility of a first order phase transition. 
\\
{\bf 3}. Continuum behavior is supported by evidence for
scaling \cite{BaSm95}. This can
be seen in the observable
\be
N'(r) = \langle \sum_y \dl_{d_{xy},r} \rangle,
\ee
the volume at geodesic distance $r$. It is maximal at some
distance $r_m$, which can be used to set a distance scale. Plotting $r_m
N'(r)/N$ for various $N$ and
suitably adjusted $\kp_2$ we find that it scales
approximately
in the sense that it approaches a
continuous function $\rh(r/r_m,\ta)$ as $N$ increases. Here
$\ta$ distinguishes
different shapes of $\rh$; for given $N$ we may think of
$\ta = \kp_2$. Hence, measuring distances in units of $r_m$  
we can let the lattice distance go to zero as $N \ra \infty$, while
keeping the shape of $\rh$ fixed. This scaling analysis needs to be
redone for larger lattices, especially in the light of the first order
nature of the phase transition.\\
{\bf 4}.  `Physical' scalar curvature observables can be obtained from
$N'(r)$: an average curvature $R_V$ and a `running' curvature 
$R_{\rm eff}$ \cite{BaSm95}.
These are negative in the crumpled phase, positive at the transition,
and appear to be ill defined in the elongated phase. This
can be compared with
predictions from (\ref{effact}) as follows. Assume that for
slow variations the
$R/G$ term in (\ref{effact}) dominates and that for
intermediate distances
spacetimes are homogenous and isotropic on the average, as
described by a
euclidean FRW scale factor $a(r)$, with effective action
\[
\frac{S_{\rm eff}}{2\pi^2} =
\int_{r_1}^{r_2} dr\, a^3 \left[\mu
 -\frac{1}{16\pi G}\, \left( \frac{\dot{a}^2}{a^2} +
\frac{1}{a^2}\right) \right].
\]
Identifying $v_{\rm eff} N'(r)$ with $2\pi^2 a(r)^3$, where
$v_{\rm eff}$ is an
effective volume, this becomes an effective action for
$N'(r)$.
Here $r_1$ should not be too small for the higher order
terms in (\ref{effact})
to be neglected, and $r_2$ should not be too large to avoid
strong fluctuations at large distances. 
We get the following stationary points:
\be
a(r) = r_0 \sinh \frac{r}{r_0},\;\;\; R = -
\frac{12}{r_0^2} = -32 \pi |G| \mu,
\label{negcur}
\ee
for $G < 0$ and
\be
a(r) = r_0 \sin  \frac{r}{r_0},\;\;\; R = +
\frac{12}{r_0^2} = 32\pi G\mu,
\label{poscur}
\ee
for $G > 0$, assuming $\mu >0$.
The first case corresponds to the negative curvatures $R_V$,
$R_{\rm eff}$ found
in the crumpled phase, the second corresponds to the
approximate four-sphere
behavior $N'(r) \propto \sin^3 r/r_0$ found in the
transition region \cite{BaSm95}. 

Further support for the negative curvature interpretation of the crumpled phase 
comes from probing the DT euclidean spacetimes with scalar test particles 
\cite{BaSmtbp}. Solving the
equation
\be
(\Delta + m_0^2)_{xy} G_{yz} = \dl_{xz},
\ee
on every configuration (where $\Delta_{xy}$ is the lattice
Laplace-Beltrami operator)
and averaging this over the configurations at fixed geodesic
distance $d_{xy} = r$ gives a propagator $G(r)$. We interprete this
propagator as corresponding to an average background geometry. The 
exponential fall-off of $G(r)$ at large distances determines
an effective mass $m$.
In the continuum, a space of constant negative curvature with
curvature radius $r_0$ as in (\ref{negcur}) gives a
nonzero $m$ even if $m_0 = 0$, namely $m = 3/r_0$.
We have measured $G(r)$ and $m$ in the crumpled
phase and found indeed nonzero effective masses, correlated with
the curvature radii $r_0 \equiv \sqrt{-12/R_V}$ obtained from
$N'(r)$.  \\

For the gravity interpretation of DT it is of course essential to 
exhibit its ability to attract.
We have investigated if there is a binding energy between two scalar test
particles, by comparing the two-particle mass $M$ extracted from 
$\langle G_{xy}^2\rangle$
with $2m$. The computation was done in the transition region 
$\kp_2 \approx \kp_2^c$,
because there the average spacetimes as seen through $N'(r)$ resemble most 
closely the 
classical $S^4$. In this case $m$ (interpreted as the constituent mass)  
is roughly proportional to $m_0$, although the ratio $m/m_0$ increases as $m_0$ gets smaller.
The results show that there is indeed a positive binding energy $E_b = 2m -M$.
We tried to see if the nonrelativistic formula 
$E_b = G^2 m^5/4$
could be used to define a renormalized Newton constant $G$,
but the binding energy did not seem to behave like $m^5$. The reason for
this may be strong finite size effects: using 
$G^2 = 4E_b/m^5$ for the smallest mass suggested
a Planck length $\sqrt{G}$ of only a third of the typical length scale of the
configurations, $r_m$ \cite{BaSm96}.

We now venture into some strong speculations about the transition.
In the infrared we expect the Einstein-Hilbert part
of the effective action (\ref{effact}) to dominate, 
because it has fewer derivatives.
So the unboundedness of this term is still relevant for
large volumes, $V/G^2 \gg 1$
(this ignores the nonlocality of possible AM terms \cite{AnMo92}).
We may follow ref.\ \cite{Haw78}
and introduce a conformal factor $\Om$ by
\be
g_{\mu\nu} = \Om^2 \bar g_{\mu\nu},\;\;\; ,
R = \Om^{-2} \bar R - 6 \Om^{-3} \bar \Delta \Om,
\ee
where $\bar R$ and $\bar \Delta$ are evaluated with the
metric $\bar g_{\mu\nu}$. The condition fixing $\Om$ is that
$\bar R$ is constant, independent of $x$. 
Making a partial integration the action takes the form
\bea
S_{\rm eff} &=& \intx \left[ \mu \Om^4 -\frac{12}{16\pi G}\,
\right.\nonumber\\
&&\left.
\left( \half\, \bar g_{\mu\nu} \dmu \Om \dnu \Om +
\frac{1}{12} \bar R \Om^2\right) + \cdots\right].
\label{omact}
\eea
The unboundedness is clear from the negative sign in front
of the derivative term. Performing the integration over
$\Om$ along the imaginary direction effectively changes this
sign to positive, which is proposed to cure the
unboundendness problem \cite{Haw78}. However, the same effect
is obtained by choosing $G < 0$, keeping $\Om$ and therefore
also the original metric real. If we assume that the saddle
point value $\mu_c$ is positive, then a possibly negative
$\bar R \Om^2$ term would be subdominant to the $\mu \Om^4$
term. So for negative $G$ the euclidean theory may also be
well defined in the infrared. This appears to contradict
perturbation theory where changing the sign of  $G$ does not
cure the unboundedness problem. 

Let us ignore this difficulty and continue {\em \bf very} schematically, 
concentrating
on the $\Om$ modes. We lump all the higher order terms into $\zt R^2$ and 
assume $\zt >0$ for stabilization. For small fluctations about some 
background we expect a
propagator of the schematic form $(-G^{-1} p^2 + \zt p^4)^{-1}$. For 
negative $G$
this is stable but for positive $G$ we expect condensation of some nonzero 
momentum modes. Suppose $L$ represents the size of the system, then 
$p \propto n/L$,
where $n$ is an integer. The modes with $-G^{-1} p^2 + \zt p^4 <0$ are
unstable, i.e. $|n| = 1,2, \dots, n_m$, with $n_m$ the largest positive integer
smaller than $L/\sqrt{\zt G}$.
Such a condensation of nonzero momentum modes may describe the branched
polymer behavior of the DT model in the elongated phase. 
The modes with $|n| > n_m$ remain uncondensed. For $G^{-1}$ sufficiently 
small, $1 > L/\sqrt{\zt G}$, $n_m = 0$ and there is still no condensation. 
This may correspond to the DT transition region, on the crumpled side, 
where we found $S^4$ behavior. If this schematic reasoning makes sense then the
maximum size of the $S^4$-like `universe' can only be of 
order of $\sqrt{\zt G}$, i.e. the Planck length.
 
Finally we recall the possibility that the DT theory may be
`trivial' \cite{BaSm95}. For negative $G$, writing
$\Om = \sqrt{16\pi |G|/12}\, \ph$ gives a
standard gradient term for $\ph$, a
$\ph^4$-like theory with coupling $\lm = 2\pi |G \Lm|/3$ 
(cf (\ref{cosco},\ref{omact})). 
If, because of triviality, this $\lm$ approaches zero 
as $N\ra \infty$ (while tuning $\kp_2$ such that we
stay on a scaling curve $\rh$), then 
we might still obtain a large size `universe' (of order $\Lm^{-1/2})$)
in relation to $\sqrt{|G|}$, which itself is arbitrarily large compared to the
lattice distance.

{\bf Acknowledgements}
The DT phenomenology reviewed above was obtained in collaboration with
B.V.~de Bakker.
We thank P.~Bia{\l}as, G.~'t~Hooft and M.~Visser
for useful discussions. 
This work is supported by FOM.

\end{document}